\documentstyle[amssymb,preprint,aps]{revtex}

\tightenlines

\begin{document}
\title{Elasticity-mediated self-organization and colloidal interactions of
solid spheres with tangential anchoring in a nematic liquid crystal}
\author{I. I. Smalyukh ($\thanks{%
E-mail: smalyukh@lci.kent.edu)}$) and O. D. Lavrentovich}
\address{Liquid Crystal Institute and Chemical Physics\\
Interdisciplinary Program, Kent State University, Kent, OH 44242-0001, USA}
\author{A. N. Kuzmin, A.V. Kachynski, and P. N. Prasad}
\address{The Institute for Lasers, Photonics, and Biophotonics, University\\
at Buffalo, The State University of New York, Buffalo, NY 14260-3000}
\date{\today }
\maketitle

\begin{abstract}
Using laser tweezers, we study colloidal interactions of solid microspheres
in the nematic bulk caused by elastic distortions around the particles with
strong tangential surface anchoring. The elastic interactions overcome the
Brownian motion when the interparticle separation $\overrightarrow{{\bf r}%
_{p}}$ is less than 3 particle diameters. The particles attract when the
angle $\theta $ between $\overrightarrow{{\bf r}_{p}}$ \ and the uniform
far-field director {\rm {\bf \^{n}}}$_{0}$ is between 0$^{0}$ and $\approx $%
70$^{0}$ and repel when 75$^{0}\lesssim \theta \leq 90^{0}$, which is
different from the model of quadrupolar interactions expected for large
separations. The particles aggregate in chains directed at $\approx $30$^{0}$
to {\rm {\bf \^{n}}}$_{0}$ and, at higher concentrations, form complex
kinetically trapped structures.
\end{abstract}


Interaction of particles in the bulk \cite%
{PoulinScience,RuhwandlTerentjev,MCLC-Prost,Fukuda,Stark-PhysRep} and at the
surfaces \cite{Capilar} of nematic liquid crystals (LCs)\ provide a rich
variety of new physical phenomena of interest to both fundamental and
applied science of liquid crystals and colloids. The studies recently
expanded into the area of optical manipulation of LC colloids \cite%
{Yada,Musevic,APL-tweezers} and even to the development of biological
sensors \cite{Abbot}. Anisotropy of molecular interactions at the particle
surface leads to elastic distortions of the LC director ${\rm {\bf \hat{n}}}$%
. A spherical particle immersed in LC with a uniform far-field director $%
{\rm {\bf \hat{n}}}_{0}$ creates distortions that are usually of dipolar or
quadrupolar symmetry \cite%
{PoulinScience,RuhwandlTerentjev,MCLC-Prost,Fukuda,Stark-PhysRep,Capilar}.
In the dipolar case, the interactions lead to chaining of particles along $%
{\rm {\bf \hat{n}}}_{0}$ \cite%
{PoulinScience,Fukuda,Stark-PhysRep,Yada,Musevic,APL-tweezers,PoulinPRE} and
a number of more complex structures \cite%
{Capilar,PoulinPRE,PoulinPRL-fer,JFukudaYokoyama,Langmuir}. \ Further
progress is impossible without a quantitative characterization of the
elasticity-mediated forces. However, the data are available only for the
case of dipolar interactions of particles with normal surface anchoring of $%
{\rm {\bf \hat{n}}}$ \cite{Yada,APL-tweezers,PoulinPRL-fer}.\ Tangentially
anchored spheres are of especial interest since they are expected to
interact as quadrupoles \cite{RuhwandlTerentjev,MCLC-Prost}. Their
interaction has never been characterized quantitatively, although it has
been shown that water droplets attract each other along the direction $%
\theta =30^{0}$ and eventually coalesce in the nematic LCs \cite{PoulinPRE}%
.\ Here we report on the first direct measurement of colloidal interactions
for spheres with tangential boundary conditions in the nematic LC. We employ
fluorescence confocal polarizing microscopy (FCPM) \cite{FCPMCPL} to
characterize the director distortions around the particles and optical
trapping with laser tweezers to measure the pair interaction force $%
\overrightarrow{{\bf F}}_{p}$\ as a function of the separation $%
\overrightarrow{{\bf r}_{p}}$ between their centers and the angle $\theta $
between $\overrightarrow{{\bf r}_{p}}$ and the far-field director ${\rm {\bf 
\hat{n}}}_{0}.$

A proper choice of materials is of prime importance for both FCPM and
optical tweezers experiments. We use a low birefringence ($\Delta n=0.04$)
nematic ZLI2806 (EM Chemicals) doped with 0.01wt.\% \ of fluorescent dye
n,n'-bis(2,5-di-tert-butylphenyl)-3,4,9,10-perylenedicarboximide (BTBP,
Aldrich) for FCPM \cite{FCPMCPL}. We use fluorescently-labeled Melamine
Resin (MR, Aldrich) spheres of diameter $D=3\mu m$. The refractive index of
MR $\ n=1.68$ is higher than the average refractive index $\overline{n}=%
\sqrt{(2n_{0}^{2}+n_{e}^{2})/3}\approx 1.49$ of ZLI2806 ($n_{0}$ is the
ordinary and $n_{e}$ is the extraordinary refractive indices) to assure
stable particle trapping \cite{APL-tweezers}. The beads were coated with
thin polyisoprene layers to produce tangentially degenerate alignment with a
vanishing azimuthal anchoring coefficient $W_{a}\lessapprox 10^{-10}J/m^{2}$
and polar anchoring coefficient $W_{p}=\left( 0.6\pm 0.4\right) \times
10^{-4}J/m^{2}$ \cite{dislocations-ch}. The cells are formed by thin (0.15
mm) glass plates; the cell gap $h=(30-100)\mu m$ is set by mylar spacers.
Thin polyimide layers (PI2555, HD MicroSystems) at the inner surfaces of the
plates are rubbed to set a uniform in-plane director ${\rm {\bf \hat{n}}}%
_{0} $.

We use a fast version of FCPM based on a Nipkow-disk confocal system
integrated with the Nikon microscope Eclipse E-600. We determine the
director configurations (through polarization-dependent fluorescence of
BTBP) as well as the positions of particles (through fluorescence of
Rhodamine B labelling the MR spheres) \cite{APL-tweezers}. We use an optical
manipulator (Solar-TII, LM-2) and a laser ($\lambda =1064nm$, used power $%
P=0-200mW$) for the dual-beam laser trapping \cite{APL-tweezers}. The
optical trap is formed by a $100\times $ objective ($NA=1.3$) \cite%
{PrasadBook}. Quantitative measurements of forces in LCs require some
additional care for three reasons: (1) the difference in refractive indices
of the particle and the host depends on the local ${\rm {\bf \hat{n}}}$, (2)
the focused light beam can reorient ${\rm {\bf \hat{n}}}$ \cite{Musevic},
and (3) light defocusing in a birefringent medium can widen the laser trap %
\cite{APL-tweezers}. We mitigated all these problems by using a
low-birefringent LC and colloidal particles larger than the waist of the
laser beam ($\approx 0.8\mu m$) \cite{APL-tweezers}.

Isolated spheres with tangential anchoring create distortions of the
quadrupolar type that quickly decay with distance $r$ from the center of the
particle and become optically undetectable at $r\gtrapprox 2D$, Fig.1a. The
director tangential to the sphere forms two point defects-boojums at the
poles of particles \cite{SoftMatterbook}. When the beads are free to move
around, they attract and form chains oriented at $\theta \approx \pm
(25^{\circ }-35^{\circ })$ with respect to ${\rm {\bf \hat{n}}}_{0}$,
Fig.1b,c. The chains elongate with time as other beads join. To verify
whether the result is not an artefact of the possible tilt of the chain with
respect to the cell plane, we used FCPM observations of the vertical
cross-sections of thick samples with chains well separated from the bounding
plates, Fig.1d. \ FCPM shows that the aggregation angle is indeed $\theta
\approx \pm (25^{\circ }-35^{\circ })$, with most of the particles joined at 
$\theta \approx \pm 30^{\circ }$. The same angle is observed not only in
ZLI2806 with nearly equal splay $K_{1}$ and bend $K_{3}$ elastic constants,
but also in pentylcyanobiphenyl (5CB) with $K_{3}/K_{1}\approx 1.56$, in the
mixtures E7 with $K_{3}/K_{1}\approx 1.5$ and ZLI3412 with $%
K_{3}/K_{1}\approx 1.1$; droplet attraction at $\theta \approx 30^{\circ }$
has been also reported by Poulin and Weitz \cite{PoulinPRE}.

The chains in the bulk can glide on conical surfaces with the apex angle $%
2\theta \approx 60^{\circ }$, Fig.1e,f. If the concentration of particles is
sufficient, the chains interact and form more complex structures, Fig.1e,f.
Chain aggregation results in kinetic trapping (jamming) of particles,
similar to the case of isotropic colloids \cite%
{JammingNature,Arest-repulsion} but different in the sense that the\
structures in LC reflect the anisotropy of interactions, Fig.1e. The
anisotropic features of the formed structures persist in a broad range of
concentrations, Fig.1b,e. Unless the sample is heated to the isotropic
phase, the jammed structures do not break apart for months.

To characterize the pair interaction, we used a dual beam laser trap. Two
particles are trapped at different locations and then released by switching
off the laser.\ Their motion is strongly influenced by Brownian motion when $%
r_{p}$ is large (pair \#1 in Fig.2a), but the anisotropic colloidal
interactions become noticeable at $r_{p}\lesssim (3-4)D$. The particles
attract if ${\bf r}_{p}$ makes the angle $\ 0^{\circ }\leq \theta
\lessapprox 70^{\circ }$ with ${\rm {\bf \hat{n}}}_{0}$ and repel if $%
75^{\circ }\lessapprox \theta \leq 90^{\circ }$ (pair ''2''\ in Fig.2a). The
attracting particles eventually touch at $\theta \approx 30^{\circ }$; the
particles in the $\theta $-range of repulsion continue to undergo
fluctuations until they enter the zone of attraction. We further probe the
pair interactions by releasing only one bead from the trap while keeping the
second trapped; the original separation is kept constant, $r_{p}=7.5\mu m$,
Fig.2b. The drift of the particles reflects the direction of interaction
force. \ 

To get a better insight into the pattern of repulsive/attractive forces, we
perform an experiment in which the position of one particle is fixed by a
strong trap ($P=100mW$) and the second particle is slowly ($\sim \mu m/s$)
moved around it by a low-intensity trap ($P_{lit}<5mW$), Fig.3. \ The second
trap follows a circle of radius $r_{t}$ centered at the first trap. The
colloidal interactions make the actual separation $r_{p}$ of the particles
different from $r_{t}$: $r_{p}<r_{t}$ when the particles attract and $%
r_{p}>r_{t}$ when they repel. \ For relatively large distances, $r_{t}>2D$,
the strongest attraction is observed at $\theta =35^{\circ }-45^{\circ }$,
Fig.2a,3, which is higher than the aggregation angle $\theta \approx
30^{\circ }$.

To determine $\overrightarrow{{\bf F}}_{p}$ we first calibrate stiffness $%
\alpha $ of the optical trap for different laser powers $P$ \cite%
{LtweezersReview}. We measure the displacement $\Delta l_{d}$ of the trapped
bead from its equilibrium position in response to the viscous drag force $%
F_{d}=3\pi D\eta _{\parallel eff}V$ as the particle is moved at a constant
speed $\overrightarrow{V}\parallel {\rm {\bf \hat{n}}}_{0}$ in the LC with
effective viscosity $\eta _{\parallel eff}$ $=5.9\cdot 10^{-5}Pa\cdot s$ \ %
\cite{Stokes}; the trap stiffness is $\alpha (P)=F_{d}/\Delta l_{d}(P)$. The
amplitude of $\overrightarrow{{\bf F}}_{p}$ is then determined by the
displacement $\Delta l_{p}$ caused by the pair interaction, $F_{p}=\alpha
\Delta l_{p}$ \cite{LtweezersReview}. \ We also use the technique of
particle escape to probe interactions at $r_{p}>1.5D$ \cite%
{Yada,APL-tweezers}. \ A single particle is moved by the tweezers with an
increasing velocity $\overrightarrow{V}\parallel {\rm {\bf \hat{n}}}_{0}$
until it escapes from the trap at which moment the ''trap escape''\ force is
determined by the Stoke's law, $F_{te}=3\pi D\eta _{\parallel eff}V$. $%
F_{te} $ is found to be a linear function of $P$, indicating that nonlinear
light-induced processes (such as realignment of ${\rm {\bf \hat{n}}}$ by the
laser beam \cite{Musevic}) are insignificant \cite{PrasadBook}. $F_{p}$ is
measured as $F_{te}$ when the escape from the trap is caused by the pair
interaction \cite{Yada,APL-tweezers}. The results obtained by the two
methods above are in a good agreement. The advantage of both approaches is
that the effects of anisotropic viscosity of LC are avoided (the velocities
of the beads during measurement of $F_{p}$ are negligibly small) and that
both $\alpha $ and $F_{te}$ are practically independent of ${\rm {\bf \hat{n}%
}}$ around the beads ($\Delta n$ is small).

To collect the data in Fig.4, the particles are placed at positions with
different $\theta $'s but constant $r_{p}$, and the force is determined by
measuring $\Delta l_{p}$. When the laser power $P$ is reduced, the direction
of bead displacement (determined from the optical microscope images)
indicates the direction of $\overrightarrow{{\bf F}}_{p}$. \ Figure 4b shows
how the angle $\gamma $ between $\overrightarrow{{\bf r}}_{p}$ and $%
\overrightarrow{{\bf F}}_{p}$ changes when one particle circumnavigates the
other. We measured the distance dependence of $\overrightarrow{{\bf F}}%
_{p}(r_{p})$ for $\theta =30^{\circ }=const$, Fig.5. \ At $r_{p}\gtrapprox
1.5D,$ the dependence $F_{p}(r_{p})$ is close to a power law $\varpropto
r_{p}^{-6}$, i.e., it is much sharper than in the case of spheres with
dipolar distortions of ${\rm {\bf \hat{n}}}$ \cite%
{Stark-PhysRep,PoulinPRL-fer}. \ The force amplitude is of the order of $K$
for $r_{p}<2D$ ($K\approx 15pN$ for ZLI2806). \ In addition to the
experiments with $D=3\mu m,$ we measured the forces for larger particles, $%
D=(4-7.5)\mu m$. \ The qualitative picture remains the same, but $%
\overrightarrow{{\bf F}}_{p}$ increases with $D$; for example, $%
\overrightarrow{{\bf F}}_{p}$ increases by $\approx 40\%$ when $D$ increases
from $3\mu m$ to $4\mu m$.

Let us compare the experimental $\overrightarrow{{\bf F}}_{p}(\theta ,r_{p})$
to available theoretical models for quadrupolar interactions with the pair
potential derived in the approximation $K_{1}=K_{3}=K$ \cite%
{RuhwandlTerentjev}:

\begin{equation}
U_{q}(r_{p},\theta )=\frac{\pi W_{p}^{2}R^{8}}{30Kr_{p}^{5}}\left( 1-\frac{%
W_{p}R}{56K}\right) \left( 9+20\cos 2\theta +35\cos 4\theta \right) ,
\label{Terentjev-Potential}
\end{equation}%
where $R=D/2$. In Fig.4 we plot the theoretical predictions for $%
\overrightarrow{{\bf F}}_{p}(\theta ,r_{p}=1.5D)$ and the $\theta $%
-dependence of the angle $\gamma $ between $\overrightarrow{{\bf r}}_{p}$
and $\overrightarrow{{\bf F}}_{p}=-\nabla U_{q}$ as calculated using\ Eq.(%
\ref{Terentjev-Potential}) with $K=15pN$ and $W_{p}=5\cdot 10^{-5}J/m^{2}$;
the calculated force is of the same order as in the experiment. The model
predicts maximum attraction at $\theta \approx 49^{\circ }$ which is
independent of $r_{p},$ while the experiment shows \ maximum attraction at $%
\theta \approx 30^{\circ }$ for $D<r_{p}<1.7D$ and at $\theta \approx
35^{\circ }-45^{\circ }$ for $2D<r_{p}<4D$. \ Furthermore, the experimental
repulsive sector is limited to $75^{\circ }\lesssim \theta \leq 90^{\circ }$%
, while the model predicts repulsion also for $\theta \approx 0$. The model
dependence $F_{p}\varpropto r_{p}^{-6}$ holds only for $r_{p}>1.5D$; at
smaller distances it is much weaker, Fig.5. \ The discrepancies are not
surprising, as the model \cite{RuhwandlTerentjev,MCLC-Prost} assumes $%
r_{p}>>D$ and weak surface anchoring, $K/W_{p}>>D.$ In the experiment,\ the
elasticity-mediated forces overcome the background Brownian motion only when 
$r_{p}<(3-4)D$; besides, $K/W_{p}\approx 0.2\mu m$, smaller than $D$. \ Note
that as $r_{p}/D$ increases, the agreement between the experiment and the
model \cite{RuhwandlTerentjev,MCLC-Prost} improves, Figs.2-5. \ To describe
the interactions at close separation, the theory should take into account
finite surface anchoring and deviations of the director from the quadrupolar
to lower symmetries, similar to the case revealed in simulations of
antiparallel elastic dipoles \cite{Fukuda}. One possible reason for the
lower symmetry of director distortions might be non-sphericity of the
particle shape and/or non-homogeneity of the surface anchoring, even when
the particle is perfectly spherical, see, for example Ref. \cite{LevYokoyama}%
.

To conclude, we characterized the pair interaction of colloidal spheres in
the nematic host that induce quadrupolar director distortions when isolated.
The interactions deviate from the quadrupolar model because the experimental
forces are significant only for small inter-particle distances. When the
concentration of particles increases, they aggregate, first in chains tilted
with respect to the far-field director and then into more complex
aggregates. The chains break apart once the LC is heated to the isotropic
phase (provided that the beads are charge-stabilized); the system,
therefore, is of interest to study kinetic arrest and jamming \cite%
{JammingNature} in anisotropic media with both attractive and repulsive
interactions \cite{Arest-repulsion}.

We thank B. Senyuk, S. Shiyanovskii, and H. Stark for discussions. We
acknowledge support of the NSF Grant DMR-0315523 (Kent) and the AFOSR DURINT
grant F496200110358 (SUNY). I.I.S. acknowledges the support of the
Fellowship of the Institute of Complex and Adaptive Matter. 

FIG.1. Colloidal aggregation in the nematic bulk: (a) FCPM texture of the
director distortions around a pair of particles; (b) FCPM image of director
distortions in the LC co-localized with the fluorescence signal from
particles (green) forming chains at $\theta \approx 30^{\circ }$ to ${\rm 
{\bf \hat{n}}}_{0}$; (c) scheme of the director distortions around the
spheres in chains; (d) FCPM\ vertical cross-sections along the d-d line in
(b) illustrating that the particle chains (green, fluorescence from
Rhodamine B) are located in the bulk of LC (gray-scale, BTBP); (e)
aggregation of spheres leading to kinetically-trapped network that reflects
anisotropy of the host even at high particle concentration $\approx 5$ wt$%
.\% $; (f) scheme of chain orientatations. The FCPM polarization is
perpendicular to ${\rm {\bf \hat{n}}}_{0}$ in (a,b) and along the chain in
(d).

FIG.2. Trajectories of particle pairs in the LC: (a) both particles are
released from the optical traps at different locations (labelled by pairs of
numbers 1 through 8; the trajectories of the two particles released at the
same time are marked by the same color); (b) one particle is fixed and the
other is released from a semicircle of radius $r_{p}\approx 7.5\mu m$ at
different $\theta $.

Fig.3. Angular dependence of the center-to-center distance $r_{p}$ between
two particles measured when one trap moves around the other trap along a
circle of a radius $r_{t}$.

FIG.4. Angular dependencies of $\overrightarrow{{\bf F}}_{p}$ for two MR
particles in the LC: (a) $F_{p}{\bf (\theta )}$ measured for $r_{p}=1.5D$
(triangles), $r_{p}=1.75D$ (squares), and $r_{p}=2D$ (circles); (b) the
angle $\gamma $ between $\overrightarrow{{\bf F}}_{p}(\theta )$ and $%
\overrightarrow{{\bf r}_{p}}$ as a function of $\theta $; (c) vector
representation of $\overrightarrow{{\bf F}}_{p}$ as determined
experimentally for $r_{p}=1.5D$ and $r_{p}=2.5D$ and calculated for $%
r_{p}\gg D$ using (\ref{Terentjev-Potential}).

FIG.5. $F_{p}$ vs. $r_{p}$ measured for $\theta \approx 30^{\circ }.$ The
solid lines are plots of $F_{p}=148(D/r_{p})^{6}pN$. The inset is the
log-log plot of the same data.


\begin{references}
\bibitem{Stokes} $\eta _{\parallel eff}$ is determined from the
Stokes-Einstein relation $\eta _{\parallel eff}=k_{b}T/3\pi \delta
_{\parallel }D$, where $k_{b}$ is the Boltzmann constant, $T$ is
temperature, and $\delta _{\parallel }$ is the diffusion coefficient
measured by analyzing Brownian motion of the bead, as described in Ref.\cite%
{Loudet2004}.
\end{references}
\end{document}